\documentclass[11pt,italian,a4widepaper]{article}

= 0.3 in \headheight = 0.2 in \oddsidemargin = 0.08 in
\usepackage{fancyhdr}            
\fancypagestyle{plain}{%
  \fancyhead{}                                     
  \fancyfoot[CE,CO]{\thepage}       
  \fancyhead[CE,CO]{\textcolor[rgb]{0.64,0.15,0.15}{Published in \emph{International Journal of Solids and Structures} \textbf{85-86}
  (2016), 76-88
\\
doi: http://dx.doi.org/10.1016/j.ijsolstr.2016.01.026}}
\setlength{\headheight}{14.5pt}}

\usepackage{latexsym}
\usepackage{amsmath,amsfonts,amssymb}
\usepackage{bm}
\usepackage{graphicx}
\usepackage{tabularx}
\usepackage{enumerate}
\usepackage{multirow}
\usepackage{booktabs}
\usepackage[font=footnotesize,labelfont=bf]{caption}
\usepackage{subfig}
\usepackage[svgnames,table]{xcolor}
\usepackage{latexsym}
\usepackage{amsmath}
\usepackage{booktabs}
\usepackage{multirow}
\usepackage{setspace}
\usepackage[toc,page]{appendix}
\usepackage{mathtools}
\usepackage{cancel}
\usepackage{relsize}
\usepackage{empheq}
\usepackage{mathrsfs}
\usepackage{breqn}
\usepackage{arcs}
\usepackage{amssymb}
\usepackage{wasysym}
\usepackage{pifont}

\begin{document}

\newcolumntype{L}[1]{>{\raggedright\let\newline\\\arraybackslash\hspace{0pt}}m{#1}}
\newcolumntype{C}[1]{>{\centering\let\newline\\\arraybackslash\hspace{0pt}}m{#1}}
\newcolumntype{R}[1]{>{\raggedleft\let\newline\\\arraybackslash\hspace{0pt}}m{#1}}

\def\ds{\displaystyle}

\newcommand{\beq}{\begin{equation}}
\newcommand{\eeq}{\end{equation}}
\newcommand{\lb}{\label}
\newcommand{\beqar}{\begin{eqnarray}}
\newcommand{\eeqar}{\end{eqnarray}}
\newcommand{\barr}{\begin{array}}
\newcommand{\earr}{\end{array}}
\newcommand{\jump}{\parallel}

\def\c{{\circ}}

\newcommand{\Ehat}{\hat{E}}
\newcommand{\That}{\hat{\bf T}}
\newcommand{\Ahat}{\hat{A}}
\newcommand{\chat}{\hat{c}}
\newcommand{\shat}{\hat{s}}
\newcommand{\khat}{\hat{k}}
\newcommand{\muhat}{\hat{\mu}}
\newcommand{\mc}{M^{\scriptscriptstyle C}}
\newcommand{\mei}{M^{\scriptscriptstyle M,EI}}
\newcommand{\mec}{M^{\scriptscriptstyle M,EC}}
\newcommand{\hbeta}{{\hat{\beta}}}
\newcommand{\rec}[2]{\left( #1 #2 \ds{\frac{1}{#1}}\right)}
\newcommand{\rep}[2]{\left( {#1}^2 #2 \ds{\frac{1}{{#1}^2}}\right)}
\newcommand{\derp}[2]{\ds{\frac {\partial #1}{\partial #2}}}
\newcommand{\derpn}[3]{\ds{\frac {\partial^{#3}#1}{\partial #2^{#3}}}}
\newcommand{\dert}[2]{\ds{\frac {d #1}{d #2}}}
\newcommand{\dertn}[3]{\ds{\frac {d^{#3} #1}{d #2^{#3}}}}

\def\bob{{\, \underline{\overline{\otimes}} \,}}
\def\ob{{\, \underline{\otimes} \,}}
\def\scalp{\mbox{\boldmath$\, \cdot \, $}}
\def\gdp{\makebox{\raisebox{-.215ex}{$\Box$}\hspace{-.778em}$\times$}}
\def\daa{\makebox{\raisebox{-.050ex}{$-$}\hspace{-.550em}$: ~$}}
\def\mK{\mbox{${\mathcal{K}}$}}
\def\cK{\mbox{${\mathbb {K}}$}}

\DeclarePairedDelimiter{\abso}{\lvert}{\rvert}
\DeclarePairedDelimiter{\norma}{\lVert}{\rVert}

\def\Xint#1{\mathchoice
   {\XXint\displaystyle\textstyle{#1}}%
   {\XXint\textstyle\scriptstyle{#1}}%
   {\XXint\scriptstyle\scriptscriptstyle{#1}}%
   {\XXint\scriptscriptstyle\scriptscriptstyle{#1}}%
   \!\int}
\def\XXint#1#2#3{{\setbox0=\hbox{$#1{#2#3}{\int}$}
     \vcenter{\hbox{$#2#3$}}\kern-.5\wd0}}
\def\ddashint{\Xint=}
\def\fpint{\Xint=}
\def\dashint{\Xint-}
\def\cpvint{\Xint-}
\def\intl{\int\limits}
\def\cpvintl{\cpvint\limits}
\def\fpintl{\fpint\limits}
\def\ointl{\oint\limits}
\def\bA{{\bf A}}
\def\ba{{\bf a}}
\def\bB{{\bf B}}
\def\bb{{\bf b}}
\def\bc{{\bf c}}
\def\bC{{\bf C}}
\def\bD{{\bf D}}
\def\bE{{\bf E}}
\def\be{{\bf e}}
\def\bbf{{\bf f}}
\def\bF{{\bf F}}
\def\bG{{\bf G}}
\def\bg{{\bf g}}
\def\bi{{\bf i}}
\def\bH{{\bf H}}
\def\bK{{\bf K}}
\def\bL{{\bf L}}
\def\bM{{\bf M}}
\def\bN{{\bf N}}
\def\bn{{\bf n}}
\def\b0{{\bf 0}}
\def\bo{{\bf o}}
\def\bX{{\bf X}}
\def\bx{{\bf x}}
\def\bP{{\bf P}}
\def\bp{{\bf p}}
\def\bQ{{\bf Q}}
\def\bq{{\bf q}}
\def\bR{{\bf R}}
\def\bS{{\bf S}}
\def\bs{{\bf s}}
\def\bT{{\bf T}}
\def\bt{{\bf t}}
\def\bU{{\bf U}}
\def\bu{{\bf u}}
\def\bv{{\bf v}}
\def\bw{{\bf w}}
\def\bW{{\bf W}}
\def\by{{\bf y}}
\def\bz{{\bf z}}
\def\T{{\bf T}}
\def\Te{\textrm{T}}
\def\Id{{\bf I}}
\def\bxi{\mbox{\boldmath${\xi}$}}
\def\balpha{\mbox{\boldmath${\alpha}$}}
\def\bbeta{\mbox{\boldmath${\beta}$}}
\def\bepsilon{\mbox{\boldmath${\epsilon}$}}
\def\bvarepsilon{\mbox{\boldmath${\varepsilon}$}}
\def\bomega{\mbox{\boldmath${\omega}$}}
\def\bphi{\mbox{\boldmath${\phi}$}}
\def\bsigma{\mbox{\boldmath${\sigma}$}}
\def\bfeta{\mbox{\boldmath${\eta}$}}
\def\bDelta{\mbox{\boldmath${\Delta}$}}
\def\btau{\mbox{\boldmath $\tau$}}
\def\tr{{\rm tr}}
\def\dev{{\rm dev}}
\def\div{{\rm div}}
\def\Div{{\rm Div}}
\def\Grad{{\rm Grad}}
\def\grad{{\rm grad}}
\def\Lin{{\rm Lin}}
\def\Sym{{\rm Sym}}
\def\Skw{{\rm Skew}}
\def\abs{{\rm abs}}
\def\Re{{\rm Re}}
\def\Im{{\rm Im}}
\def\capB{\mbox{\boldmath${\mathsf B}$}}
\def\capC{\mbox{\boldmath${\mathsf C}$}}
\def\capD{\mbox{\boldmath${\mathsf D}$}}
\def\capE{\mbox{\boldmath${\mathsf E}$}}
\def\capG{\mbox{\boldmath${\mathsf G}$}}
\def\tcapG{\tilde{\capG}}
\def\capH{\mbox{\boldmath${\mathsf H}$}}
\def\capK{\mbox{\boldmath${\mathsf K}$}}
\def\capL{\mbox{\boldmath${\mathsf L}$}}
\def\capM{\mbox{\boldmath${\mathsf M}$}}
\def\capR{\mbox{\boldmath${\mathsf R}$}}
\def\capW{\mbox{\boldmath${\mathsf W}$}}

\def\i{\mbox{${\mathrm i}$}}
\def\mC{\mbox{\boldmath${\mathcal C}$}}
\def\mB{\mbox{${\mathcal B}$}}
\def\mE{\mbox{${\mathcal{E}}$}}
\def\mL{\mbox{${\mathcal{L}}$}}
\def\mK{\mbox{${\mathcal{K}}$}}
\def\mV{\mbox{${\mathcal{V}}$}}
\def\C{\mbox{\boldmath${\mathcal C}$}}
\def\E{\mbox{\boldmath${\mathcal E}$}}

\def\ACME{{ Arch. Comput. Meth. Engng.\ }}
\def\ARMA{{ Arch. Rat. Mech. Analysis\ }}
\def\AMR{{ Appl. Mech. Rev.\ }}
\def\ASCEEM{{ ASCE J. Eng. Mech.\ }}
\def\acta{{ Acta Mater. \ }}
\def\CMAME {{ Comput. Meth. Appl. Mech. Engrg.\ }}
\def\CRAS{{ C. R. Acad. Sci., Paris\ }}
\def\EFM{{ Eng. Fract. Mech.\ }}
\def\EJMA{{ Eur.~J.~Mechanics-A/Solids\ }}
\def\IJES{{ Int. J. Eng. Sci.\ }}
\def\IJF{{ Int. J. Fracture}}
\def\IJMS{{ Int. J. Mech. Sci.\ }}
\def\IJNAMG{{ Int. J. Numer. Anal. Meth. Geomech.\ }}
\def\IJP{{ Int. J. Plasticity\ }}
\def\IJSS{{ Int. J. Solids Structures\ }}
\def\IngA{{ Ing. Archiv\ }}
\def\JAM{{ J. Appl. Mech.\ }}
\def\JAP{{ J. Appl. Phys.\ }}
\def\JE{{ J. Elasticity\ }}
\def\JM{{ J. de M\'ecanique\ }}
\def\JMPS{{ J. Mech. Phys. Solids\ }}
\def\Macro{{ Macromolecules\ }}
\def\MOM{{ Mech. Materials\ }}
\def\MMS{{ Math. Mech. Solids\ }}
\def\MMT{{ Metall. Mater. Trans. A}}
\def\MPCPS{{ Math. Proc. Camb. Phil. Soc.\ }}
\def\MSE{{ Mater. Sci. Eng.}}
\def\PMPS{{ Proc. Math. Phys. Soc.\ }}
\def\PRE{{ Phys. Rev. E\ }}
\def\PRSL{{ Proc. R. Soc.\ }}
\def\PRL{{ Phys. Rev. Letters\ }}
\def\rock{{ Rock Mech. and Rock Eng.\ }}
\def\QAM{{ Quart. Appl. Math.\ }}
\def\QJMAM{{ Quart. J. Mech. Appl. Math.\ }}
\def\SCRMAT{{ Scripta Mater.\ }}
\def\SM{{\it Scripta Metall. }}

\def\salto#1#2{
[\mbox{\hspace{-#1em}}[#2]\mbox{\hspace{-#1em}}]}

\title{Isotoxal star-shaped polygonal voids and rigid inclusions in nonuniform antiplane shear fields.\\
Part II:  Singularities, annihilation and invisibility}\date{}

\author{F. Dal Corso, S. Shahzad and D. Bigoni \\
DICAM, University of Trento, via Mesiano 77, I-38123 Trento, Italy }

\maketitle

\begin{abstract}
\noindent
Notch stress intensity factors and stress intensity factors are obtained analytically for isotoxal star-shaped polygonal voids and rigid inclusions (and also for the corresponding limit cases of star-shaped cracks and stiffeners), when loaded through remote inhomogeneous (self-equilibrated, polynomial) antiplane shear stress in an infinite linear elastic matrix.
Usually these solutions show stress singularities at the inclusion corners. It is shown that an infinite set of
geometries and loading conditions exist for which not only the singularity is absent, but the stress vanishes (\lq annihilates')  at the corners.
Thus the material, which even without the inclusion corners would have a finite stress, remains unstressed at these points
in spite of the applied remote load.
Moreover, similar conditions are determined in which a star-shaped crack or stiffener leaves the ambient stress completely unperturbed, thus reaching a condition
of \lq quasi-static invisibility'.
Stress annihilation and invisibility define optimal loading modes for the overall strength of a composite
 and are useful for designing ultra-resistant materials.
\end{abstract}


\section{Introduction}

The knowledge of the stress intensity factor (SIF) and of the notch stress intensity factor (NSIF), respectively, for star-shaped cracks/stiffeners and isotoxal star-shaped polygonal voids/rigid-inclusions
is crucial as they represent failure criteria in the design of brittle-matrix composites \cite{anderson}. Therefore, results presented in Part I \cite{partI} of this study are complemented with the analytical, closed-form determination of SIF and NSIF.
In this way, a full characterization of the stress fields near star-shaped cracks/stiffeners and polygonal voids/rigid-inclusions is reached.
This allows for the analysis of the conditions of inclusion neutrality that occur
 when the ambient field is left unperturbed outside the inclusion. The neutrality condition has been thoroughly
 analyzed \cite{benveniste, serkov, mahboob, ru, rus, vasudevan, vasudevan2, wang} because it provides a criterion
 for the introduction of an inclusion in a composite without a loss of strength and because
 it is a problem linked to homogenization techniques for composites \cite{christensen}.

Recently, elastic cloaking in metamaterials has demonstrated wave invisibility \cite {brun1, brun2, colquit, farath, milton, misseroni, norris}
and is, in a sense, a dynamic counterpart to neutrality.
Both neutrality and invisibility are strong conditions that cannot be achieved in an exact sense for a perfectly bonded inclusion \cite{brun2, ru}.
So neutrality, in statics, has been relaxed with the introduction of \lq quasi-neutrality' \cite{bertoldi}, which
allows rapidly decaying stress singularities at the inclusion boundary to be neglected.
In fact, considering a problem of inclusion involving singularities
(as for instance at an inclusion vertex) it seems at a first glance impossible to achieve neutrality or invisibility. Nevertheless,
it will be demonstrated in this article that
two special cases exist for an infinite class of geometries and modes of loading in which the \lq usual' stress singularity is absent.
One of  these situations
occurs at the vertex of a star-shaped void or rigid inclusion and
will be termed \lq stress annihilation', while the other, occurring for a star-shaped crack or stiffener, will be
termed \lq quasi-static invisibility' (or full neutrality).
In the former case, the stress {\it vanishes} at the corner of the void/inclusion, instead of displaying the singularity which would be usually expected at a sharp corner, while, in the
latter case, the star-shaped crack or star-shaped inclusion leaves the ambient stress field {\it completely unperturbed}, so that the inclusion becomes \lq invisible' or \lq fully neutral' (the word \lq neutrality' is weak,
because in the case analyzed in this article the stress remains completely undisturbed everywhere in the matrix, so that the inclusion simply \lq disappears').
Note that the conditions of stress annihilation and invisibility imply that the material does not fail at the void/inclusion/crack/stiffener points, but far from them and only when the material would break in the unperturbed problem.
It is also shown that there are specific situations in which a partial invisibility and a partial stress annihilation is reached. In these cases invisibility or stress annihilation are verified
at some but not all of the points of the star-shaped crack/stiffener or void/inclusion,
so that in these cases failure of the material occurs at the points where the stress remains singular.

The results obtained in the present article (and in Part I) refer to regular shapes of inclusions/cracks and to an infinite elastic domain, so that it is natural to address the question on how these two idealizations affect the results, particularly for quasi-static invisibility and stress annihilation. It is therefore shown that these situations can also be met for irregular star-shaped voids and cracks. Moreover, a numerical (finite elements) analysis shows that the features are present also for finite domains.

The present article is organized as follows.
In Section \ref{SIIIFFFF} the SIFs and NSIFs, respectively for star-shaped cracks and stiffeners and for isotoxal
star-shaped void and rigid inclusions will be determined.
Note that the determination is in a closed-form, so that the solution does not involve infinite series.
In section \ref{nonnainvisibile} results will be presented in terms of stress fields around voids and cracks.
Also in Section \ref{nonnainvisibile} the conditions for quasi-static invisibility and stress annihilation will be explained in detail,
together with the situations of partial invisibility and partial stress annihilation. Generalizations to
irregular star-shaped voids/inclusions (and cracks/stiffeners) and inclusions in a finite domain will be covered in Section \ref{sonostanco}.

\section{Stress and Notch Intensity Factors} \lb{SIIIFFFF}

A measure of the stress intensification at an inclusion vertex can be obtained through the evaluation of the Stress Intensity Factor (SIF)
for a star-shaped crack or a stiffener and of the Notch Stress Intensity Factor (NSIF), in the case of a polygonal void or rigid inclusion.
The definition of these factors is given in relation to the specific form of remote shear stress ($\tau_{\rho 3}$ and $\tau_{\theta 3}$ in a polar coordinate system $\rho$, $\theta$, and $x_3$), in a way that the asymptotic singular fields
are represented by a constant depending only on the boundary conditions \cite{radaj}. In particular, with reference to the decomposition
(considered in Part I) of the displacement field $w$ in its symmetric and antisymmetric parts,
\beq
\tau_{\rho3}^{\textup{A}}(\rho,0)=\tau_{\vartheta3}^{\textup{S}}(\rho,0)=0,
\eeq
the definition of SIF is introduced for star-shaped crack or stiffener
\beq\label{sif}
K_{\textup{III}}^{\textup{S}}=
\lim_{\rho \rightarrow 0} \sqrt{2\pi \rho} \,\,
\tau_{\rho3}(\rho,0),\qquad
K_{\textup{III}}^{\textup{A}}=\lim_{\rho \rightarrow 0} \sqrt{2\pi \rho} \,\,
\tau_{\vartheta3}(\rho,0),
\eeq
while, for star-shaped void or rigid inclusion, the definition of NSIF is introduced as
\beq\label{nsif}
K_{III}^{\textup{S}}= \lim_{\rho \rightarrow 0} \sqrt{2\pi}\,\,
\rho^{-\lambda_1^{\textup{S}}} \tau_{\rho3}(\rho,0),\qquad K_{III}^{\textup{A}}=\lim_{\rho
\rightarrow 0} \sqrt{2\pi}\,\, \rho^{-\lambda_1^{\textup{A}}}
\tau_{\vartheta3}(\rho,0).
\eeq

Note that, differently from the definition of SIF, the definition of NSIF provides
non-null values even in the case of non-singular leading-order terms,
corresponding to the case of a non-negative smaller eigenvalue, $\lambda_1\geq0$ (see Table 1 in Part I of this study).

Following the procedure proposed in \cite{japan}, a closed-form expression for $K_{III}^{(m)}$ is obtained, by
considering isotoxal star-shaped voids or star-shaped cracks and rigid-inclusions within an isotropic matrix
subject to a remote generic polynomial stress condition of order $m$ (as introduced in Part I).
The SIF and NSIF, obtained with reference to the situation in which an inclusion vertex lies
on the positive part of $x_1$ axis (see Fig.$2$ in Part I), are crucial to predict fracture initiation (or propagation), via energy release rate, \cite{energy1,energy2,energy3}.

\subsection{SIF for star-shaped crack or stiffener}

Using the complex potential obtained in Part I for the case of $n$-pointed star-shaped crack or stiffener (equation (51) of Part I), the stress field  can be expressed in the
transformed plane as
\beq
\label{eq_tauconf_star_crack}
\tau_{13}^{(m)}-\i \tau_{23}^{(m)}= \frac{a^m\, 2^{\frac{2}{n}-t}}{\left(1-\zeta^{-n}\right)\left(1+\zeta^{-n}\right)^{\frac{2-n}{n}}}
\ds\sum_{j=0}^{q}  \frac{(m+1-j n)}{j!}  \left(T^{(m)}\zeta^{m-jn} -\chi \overline{T^{(m)}}\frac{1}{\zeta^{m+2-j n}}\right)\prod_{l=0}^{j-1} (t-l) ,
\eeq
where $\chi=1$ ($\chi=-1$) for a void (for a rigid inclusion).
Equation (\ref{eq_tauconf_star_crack}) can be expanded about the vertex of the inclusion (at $x_1 = a$ and $x_2=0$), by introducing $\zeta=1+\zeta^*$ with $|\zeta^*|\rightarrow 0$
(corresponding to $z=a+z^*$ with $|z^*|\rightarrow 0$, see Fig. 2 in Part I), as
\beq
 \tau_{13}^{(m)}-\i \tau_{23}^{(m)}  \simeq   \frac{a^m\,  \left[b^{(m)}_0(1-\chi)
-\i c^{(m)}_0(1+\chi)\right]}{2^{t-1} n(m+1)\zeta^*}\ds \sum_{j=0}^{q} \frac{(m+1-j n)}{j!} \prod_{l=0}^{j-1} (t-l) .
\eeq

Considering now the conformal mapping [equation (42) of Part I],
the relationship between the physical coordinate $z^*=\rho e^{\i\vartheta}$ and its conformal counterpart $\zeta^*$ can be expanded
as
\beq
\label{eq_inversa}
\zeta^* \simeq \ds 2 \sqrt{\frac{\rho}{n a}  }\,\, e^{\frac{\i\vartheta}{2}},
\eeq
so that the  following asymptotic expansion in the physical plane is obtained
\begin{dmath}
\label{asym_poly_z_star_crack}
 \tau_{13}^{(m)}-\i \tau_{23}^{(m)} \simeq  \frac{a^m\, \sqrt{a} \left[b^{(m)}_0(1-\chi)
-\i c^{(m)}_0(1+\chi)\right]}{2^{t} (m+1) \sqrt{n \rho} }
\left(\cos\frac{\vartheta}{2}-\i\, \sin\frac{\vartheta}{2}\right)\sum_{j=0}^{q} \frac{(m+1-j n)}{j!} \prod_{l=0}^{j-1} (t-l),
\end{dmath}
from which, according to the asymptotic description (Sect.$2.2$ in Part I),
the  square root stress singularity at the star point is evident.
Using definition (\ref{sif}), the SIF can be evaluated  as
\beq
\label{SIFSIF0}
\begin{Bmatrix}
K_{\textup{III}}^{\textup{S}}(n,m)\\[3mm]
K_{\textup{III}}^{\textup{A}}(n,m)
\end{Bmatrix}
=   \frac{2^{\frac{n-4(m+1)}{2n}}\,a^m}{m+1}\sqrt{\frac{\pi a}{n}}
\left[\sum_{j=0}^{q} \frac{(m+1-j n)}{j!} \prod_{l=0}^{j-1} (t-l)\right]
\begin{Bmatrix}
(1-\chi)b_0^{(m)}\\[3mm]
(1+\chi)c_0^{(m)}
\end{Bmatrix}
,
\eeq
so  that the SIFs for a star-shaped crack $K_{\textup{III}}^{\text{{\tiny\ding{73}}}}(n,m)$ or a star-shaped stiffener $K_{\textup{III}}^{\text{{\tiny\ding{72}}}}(n,m)$ are
\beq\label{eq_sif}
\begin{Bmatrix}
K_{\textup{III}}^{\text{{\tiny\ding{72}}}\textup{S}}(n,m)\\[3mm]
K_{\textup{III}}^{\text{{\tiny\ding{73}}}\textup{A}}(n,m)
\end{Bmatrix}
=   \frac{2^{\frac{3n-4(m+1)}{2n}}\,a^m}{m+1}\sqrt{\frac{\pi a}{n}}
\left[\sum_{j=0}^{q} \frac{(m+1-j n)}{j!} \prod_{l=0}^{j-1} (t-l)\right]
\begin{Bmatrix}
b_0^{(m)}\\[3mm]
c_0^{(m)}
\end{Bmatrix},
\eeq
and $K_{\textup{III}}^{\text{{\tiny\ding{72}}}\textup{A}}(n,m)=K_{\textup{III}}^{\text{{\tiny\ding{73}}}\textup{S}}(n,m)=0$.

\begin{figure}[!htb]
  \begin{center}
\includegraphics[width=12 cm]{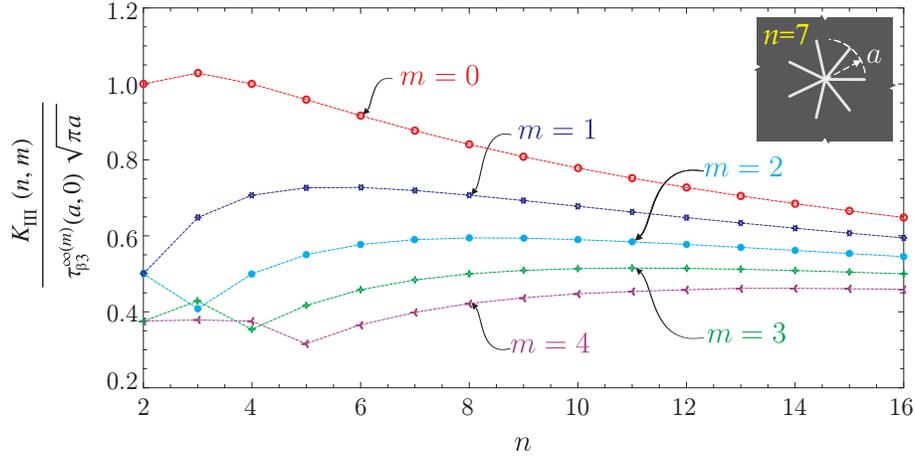}
\caption{ Stress intensity factors for both star-shaped cracks
${K}_{\textup{III}}^{\text{{\tiny\ding{73}}}}$ and star-shaped stiffeners ${K}_{\textup{III}}^{\text{{\tiny\ding{72}}}}$ as
 functions of the number $n$ of crack or stiffener tips, for different orders $m$ of the applied remote
 polynomial antiplane shear loading. Note that, due to the division
by the unperturbed stress $\tau_{\beta3}^\infty$ evaluated at the inclusion vertex ($a,0$),  cracks ($\beta=1$) and  stiffeners ($\beta=2$) display the same SIF,
independently of the loading parameters
$b_0^{(m)}$ and $c_0^{(m)}$.
A single crack or stiffener corresponds to $n=2$, which in all cases does not correspond to the maximum SIF.
}
\label{fig_part_I_sif}
 \end{center}
\end{figure}

Taking into account the definition of remote applied shear stress $\tau^\infty_{13}$ and $\tau^\infty_{23}$ [see equation ($7$) of Part I], which provides the unperturbed stress components as a function of
the loading parameters $b_0^{(m)}, c_0^{(m)}$, the Stress Intensity Factors (\ref{eq_sif}) can be rewritten as
\beq
\label{eq_sif_speciali}
\begin{Bmatrix}
K_{\textup{III}}^{\text{{\tiny\ding{72}}}\textup{S}}(n,m)\\[3mm]
K_{\textup{III}}^{\text{{\tiny\ding{73}}}\textup{A}}(n,m)
\end{Bmatrix}
=\frac{2^{\frac{3n-4(m+1)}{2n}}}{(m+1)\sqrt{n}}
\left[\sum_{j=0}^{q} \frac{(m+1-j n)}{j!} \prod_{l=0}^{j-1} (t-l)\right]
\sqrt{\pi a}
\left\{
\begin{array}{ll}
\tau_{13}^{\infty(m)}(a,0)\\
\tau_{23}^{\infty(m)}(a,0)
\end{array}
\right\}.
\eeq
The SIFs are reported in Fig. \ref{fig_part_I_sif} as functions of the number $n$ of the tips and for different orders $m$ of the applied remote polynomial
antiplane shear loading.
Expression (\ref{eq_sif_speciali}) for the SIFs simplifies in the special cases listed below.
\begin{itemize}
\item
$t=2(m+1)/n \in \mathbb{N}$,
\beq
\label{eq_sif_speciali1}
\begin{Bmatrix}
K_{\textup{III}}^{\text{{\tiny\ding{72}}}\textup{S}}(n,m)\\[3mm]
K_{\textup{III}}^{\text{{\tiny\ding{73}}}\textup{A}}(n,m)
\end{Bmatrix}
=\frac{2^{\frac{3n-4(m+1)}{2n}}}{(m+1)\sqrt{n}} \sum_{j=0}^{q} \frac{t!(m+1-j n)}{j!(t-j)!}
\sqrt{\pi a}
\begin{Bmatrix}
\tau_{13}^{\infty(m)}(a,0)\\[3mm]
\tau_{23}^{\infty(m)}(a,0)
\end{Bmatrix}
.
\eeq
\item
$n>m+1$ (and therefore $q=0$),
\beq
\label{eq_sif_speciali2}
\begin{Bmatrix}
K_{\textup{III}}^{\text{{\tiny\ding{72}}}\textup{S}}(n,m)\\[3mm]
K_{\textup{III}}^{\text{{\tiny\ding{73}}}\textup{A}}(n,m)
\end{Bmatrix}
=\frac{2^{\frac{3n-4(m+1)}{2n}}}{\sqrt{n}}  \sqrt{\pi a}
\left\{
\begin{array}{ll}
\tau_{13}^{\infty(m)}(a,0)\\[3mm]
\tau_{23}^{\infty(m)}(a,0)
\end{array}
\right\} .
\eeq

This case embraces an infinite set of solutions, one such solution is that for a cruciform crack ($n=4$, Fig.$4$ of Part I)
subject to uniform, linear and quadratic remote antiplane shear load ($m=0,1,2$).
\item
$n=2$ (crack or stiffener inclusions),
\beq
\begin{Bmatrix}
K_{\textup{III}}^{\text{{\tiny\ding{72}}}\textup{S}}(n,m)\\[3mm]
K_{\textup{III}}^{\text{{\tiny\ding{73}}}\textup{A}}(n,m)
\end{Bmatrix}
=\frac{\sqrt{\pi a}}{2^m (m+1)}
\sum_{j=0}^{q} \frac{(m+1)!(m+1-2j)}{j!(m+1-j)!}
\left\{
\begin{array}{ll}
\tau_{13}^{\infty(m)}(a,0)\\[3mm]
\tau_{23}^{\infty(m)}(a,0)
\end{array}
\right\}
.
\eeq
\end{itemize}
Note that, in the particular case of uniform antiplane shear ($m=0$), equation (\ref{eq_sif_speciali2})
provides the same result as equation (32) in \cite{sih}.

\begin{figure}[!htb]
  \begin{center}
\includegraphics[width=12 cm]{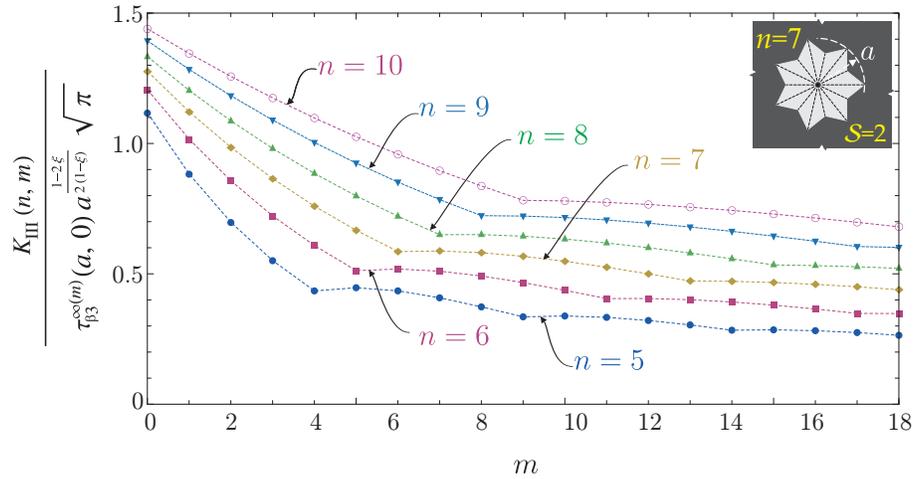}
\caption{
Notch stress intensity factors for both isotoxal star-shaped polygonal voids ${K}_{\textup{III}}^{\text{\tiny\ding{73}}}$  and rigid inclusions ${K}_{\textup{III}}^{\text{{\tiny\ding{72}}}}$ for ${\mathcal S}=2$,
functions of the order $m$ of the applied remote polynomial antiplane shear loading. Note that, due to the division
by the unperturbed stress $\tau_{\beta3}^\infty$ evaluated at the inclusion vertex ($a,0$),  voids ($\beta=1$) and  rigid inclusions ($\beta=2$) display the same NSIF,
independently of the loading parameters
$b_0^{(m)}$ and $c_0^{(m)}$.
A uniform remote shear stress field corresponds to $m=0$, which in all cases corresponds to the maximum NSIF.
}
\label{fig_part_II_sif}
 \end{center}
\end{figure}

\subsection{NSIF for isotoxal star-shaped void and rigid inclusion}

For the derivation of SIFs and NSIFs, different methods (FE analysis, BE analysis, Singular integral equations etc.)
are involved \cite{radaj,notch1,notch2,notch3,notch4,notch5}, so that closed-form expressions of SIFs and NSIFs using full-field information,
to the best of authors' knowledge, have only been provided for the case of polygonal voids and rigid inclusions inclusions
under uniform anti-plane condition ($m=0$) \cite{japan}. The NSIFs at the vertex of an $n$-pointed isotoxal star void or rigid inclusion
within an elastic
material subject to remote polynomial loading are analytically evaluated.
For the considered $n$-pointed isotoxal star polygon, the external semi-angle at each vertex is $\alpha=\pi(1-\xi)$, so assuming
that the angle parameter $\xi$ ranges within $\left[0,1/2-1/n\right]$,
the leading-order terms in the asymptotic expansion of the stress fields about a vertex are associated to the following eigenvalues
\beq
\begin{cases}
\lambda^{\text{\tiny\ding{73}}\textup{S}}_1(\xi)=\lambda^{\text{\tiny\ding{72}}\textup{A}}_1(\xi)=\dfrac{\xi}{1-\xi}\geq 0, \\[5mm]
\lambda^{\text{\tiny\ding{73}}\textup{A}}_1(\xi)=\lambda^{\text{\tiny\ding{72}}\textup{S}}_1(\xi)=-\dfrac{1-2\xi}{2(1-\xi)}<0.
\end{cases}
\eeq

Using the complex potential obtained in Part I, eqn ($72$),
and the derivative of the conformal mapping (56),
the stress field in the transformed plane can be obtained from equation ($30$)$_2$ as
\beq\label{tauconf_star_generale}
\tau_{13}^{(m)}-\i \tau_{23}^{(m)}= \frac{\left(a\,\Omega(n,\xi)\right)^{m}\zeta^{2}\ds\sum_{j=0}^{q} (m+1-j n)L_{m+1-j n}
\left(T^{(m)}\zeta^{m-jn} -\chi \overline{T^{(m)}}\frac{1}{\zeta^{m+2-j n}}\right)}{\left(\zeta^{n}-1\right)^{1-2\xi}
\left(\zeta^{n}+1\right)^{2\left(\xi+\frac{1}{n}\right)-1}},
\eeq
which
can be expanded about the neighborhood of the vertex (at $x_1 = a$ and $x_2=0$)
by introducing $\zeta=1+\zeta^*$ with $|\zeta^*|\rightarrow 0$ (corresponding to $z=a+z^*$ with $|z^*|\rightarrow 0$) as
\beq
 \tau_{13}^{(m)}-\i \tau_{23}^{(m)}  \simeq   \frac{\left(a\,\Omega(n,\xi)\right)^{m}}{m+1} \frac{\left[(1-\chi)b^{(m)}_0
-\i(1+\chi) c^{(m)}_0\right]}{n^{1-2\xi} \left(\zeta^*\right)^{1-2\xi}} 2^{1-2\xi-\frac{2}{n}} \sum_{j=0}^{q}  (m+1-j n) L_{m+1-j n}.
\eeq
A further exploitation of the first derivative of the conformal mapping
($56$) discloses the asymptotic relation
between the physical coordinate $z^*=\rho e^{\i\vartheta}$ and its conformal counterpart $\zeta^*$
\beq
\label{eq_relationship_between_two_planes_regolare_star_generale}
\zeta^* \simeq \ds
2^{\frac{n(1-\xi)-1}{n(1-\xi)}
}
n^{\frac{2\xi-1}{2(1-\xi)}}\left(\frac{1-\xi}{a \,\Omega(n,\xi)}\rho e^{\i\vartheta}\right)^{\frac{1}{2(1-\xi)}},
\eeq
which leads to the  following asymptotic expansion in the physical plane
\begin{dmath}
\begin{array}{lll}
\label{asym_poly_z_star_generale}
 \tau_{13}^{(m)}-\i \tau_{23}^{(m)}  \simeq  &\ds \frac{\left(a\,\Omega(n,\xi)\right)^{m}}{(m+1)2^{\frac{1}{n(1-\xi)}}}
  \left(\frac{a \,\Omega(n)}{n(1-\xi)\rho}\right)^{\frac{1-2\xi}{2(1-\xi)}}
\left[(1-\chi)b^{(m)}_0
-\i (1+\chi)c^{(m)}_0\right]\\[6mm]
&\ds\qquad\left[\cos\left(\dfrac{1-2\xi}{2(1-\xi)} \vartheta\right)-\i\, \sin\left(\dfrac{1-2\xi}{2(1-\xi)} \vartheta\right)\right]
\sum_{j=0}^{q}  (m+1-j n) L_{m+1-j n}.
\end{array}
\end{dmath}
From the asymptotic expansion of the stress field (\ref{asym_poly_z_star_generale}),
the NSIFs associated to the (singular or non-singular) leading-order term
can be evaluated using definition (\ref{nsif})  as
\beq
\label{eq_nsif}
\left\{
\begin{array}{ll}
K_{\textup{III}}^{\text{{\tiny\ding{72}}}\textup{S}}(n,m)\\[3mm]
K_{\textup{III}}^{\text{{\tiny\ding{73}}}\textup{A}}(n,m)
\end{array}
\right\}
=\frac{2\sqrt{2\pi} \left(a\,\Omega(n,\xi)\right)^{m}}{(m+1)2^{\frac{1}{n(1-\xi)}}}
\left(\frac{a \,\Omega(n,\xi)}{n(1-\xi)}\right)^{\frac{1-2\xi}{2(1-\xi)}}
\sum_{j=0}^{q}  (m+1-j n) L_{m+1-j n}
\left\{
\begin{array}{ll}
b_0^{(m)}\\
c_0^{(m)}
\end{array}
\right\},
\eeq
which, considering the value of the unperturbed stress components  $\tau^\infty_{13}$ and $\tau^\infty_{23}$ [see equation ($7$) of Part I] at the
void/inclusion vertex $(x_1=a,x_2=0)$, can be rewritten as
\beq\label{nsifnsif}
\left\{
\begin{array}{ll}
K_{\textup{III}}^{\text{{\tiny\ding{72}}}\textup{S}}(n,m)\\[3mm]
K_{\textup{III}}^{\text{{\tiny\ding{73}}}\textup{A}}(n,m)
\end{array}
\right\}
=\frac{2\sqrt{2\pi} \left(\Omega(n,\xi)\right)^{m}}{(m+1)2^{\frac{1}{n(1-\xi)}}}
\left(\frac{a \,\Omega(n,\xi)}{n(1-\xi)}\right)^{\frac{1-2\xi}{2(1-\xi)}}
\sum_{j=0}^{q}  (m+1-j n) L_{m+1-j n}
\left\{
\begin{array}{ll}
\tau_{13}^{\infty(m)}(a,0)\\
\tau_{23}^{\infty(m)}(a,0)
\end{array}
\right\}
.
\eeq

The Notch Stress Intensity Factors (\ref{nsifnsif}) simplify in the special cases listed below.
\begin{itemize}
\item  $n$-sided regular polygon
\beq
\left\{
\begin{array}{ll}
K_{\textup{III}}^{\text{{\tiny\ding{72}}}\textup{S}}(n,m)\\[3mm]
K_{\textup{III}}^{\text{{\tiny\ding{73}}}\textup{A}}(n,m)
\end{array}
\right\}
=\frac{2 \sqrt{2\pi} \left(\Omega(n)\right)^{m}}{m+1} \left(\frac{a \,\Omega(n)}{n+2}\right)^{\frac{2}{n+2}}
\sum_{j=0}^{q}  (m+1-j n) L_{m+1-j n}
\left\{
\begin{array}{ll}
\tau_{13}^{\infty(m)}(a,0)\\
\tau_{23}^{\infty(m)}(a,0)
\end{array}
\right\}
.
\eeq

\item  $n$-pointed regular star polygon with density $\mathcal{S}=2$ (with $n\geq 4$),

\beq
\left\{
\begin{array}{ll}
K_{\textup{III}}^{\text{{\tiny\ding{72}}}\textup{S}}(n,m)\\[3mm]
K_{\textup{III}}^{\text{{\tiny\ding{73}}}\textup{A}}(n,m)
\end{array}
\right\}
=\frac{2^{\frac{n+6}{n+4}}  \sqrt{2\pi} \left(\Omega(n)\right)^{m}}{m+1}  \left(\frac{a \,\Omega(n)}{n+4}\right)^{\frac{4}{n+4}}
\sum_{j=0}^{q}  (m+1-j n) L_{m+1-j n}
\left\{
\begin{array}{ll}
\tau_{13}^{\infty(m)}(a,0)\\
\tau_{23}^{\infty(m)}(a,0)
\end{array}
\right\}
.
\eeq

\end{itemize}

Moreover, in the case that $m+1<n$ (so that $q=0$) the following identity holds
\beq
\sum_{j=0}^{q}  (m+1-j n) L_{m+1-j n}=m+1.
\eeq
This case embraces an infinite set of solutions, one such solution is that for a
$5$-pointed isotoxal star-shaped void or rigid  inclusion subject to uniform, linear,
quadratic and cubic remote antiplane shear load ($m=0,1,2,3$). Therefore, the Notch Stress Intensity Factors (\ref{nsifnsif})
for the generic $n$-pointed isotoxal star polygon reduces to
\beq
\left\{
\begin{array}{ll}
K_{\textup{III}}^{\text{{\tiny\ding{72}}}\textup{S}}(n,m)\\[3mm]
K_{\textup{III}}^{\text{{\tiny\ding{73}}}\textup{A}}(n,m)
\end{array}
\right\}
=\frac{2 \sqrt{2\pi} \left(\Omega(n,\xi)\right)^{m}} {2^{\frac{1}{n(1-\xi)}}}
\left(\frac{a \,\Omega(n,\xi)}{n(1-\xi)}\right)^{\frac{1-2\xi}{2(1-\xi)}}
\left\{
\begin{array}{ll}
\tau_{13}^{\infty(m)}(a,0)\\[3mm]
\tau_{23}^{\infty(m)}(a,0)
\end{array}
\right\},
\eeq
which simplifies in the following special cases to
\begin{itemize}
\item  $n$-sided regular polygon
\beq
\left\{
\begin{array}{ll}
K_{\textup{III}}^{\text{{\tiny\ding{72}}}\textup{S}}(n,m)\\[3mm]
K_{\textup{III}}^{\text{{\tiny\ding{73}}}\textup{A}}(n,m)
\end{array}
\right\}
=2 \sqrt{2\pi} \left(\Omega(n)\right)^{m} \left(\frac{a \, \Omega(n)}{n+2}\right)^{\frac{2}{n+2}}
\left\{
\begin{array}{ll}
\tau_{13}^{\infty(m)}(a,0)\\[3mm]
\tau_{23}^{\infty(m)}(a,0)
\end{array}
\right\}.
\eeq

\item  $n$-pointed regular star polygon with density $\mathcal{S}=2$ (with $n\geq 4$),

\beq
\left\{
\begin{array}{ll}
K_{\textup{III}}^{\text{{\tiny\ding{72}}}\textup{S}}(n,m)\\[3mm]
K_{\textup{III}}^{\text{{\tiny\ding{73}}}\textup{A}}(n,m)
\end{array}
\right\}
=2^{\frac{n+6}{n+4}} \sqrt{2\pi}  \left(\Omega(n)\right)^{m} \left(\frac{a \, \Omega(n)}{n+4}\right)^{\frac{4}{n+4}}
\left\{
\begin{array}{ll}
\tau_{13}^{\infty(m)}(a,0)\\[3mm]
\tau_{23}^{\infty(m)}(a,0)
\end{array}
\right\}.
\eeq
\end{itemize}

\section{Stress fields, annihilation and invisibility} \lb{nonnainvisibile}

Results obtained in terms of the full-field solution (Part I of this study) allows for
the complete determination of the stress field near an isotoxal star-shaped polygonal void or rigid inclusion, including the limits of star-shaped cracks and stiffeners.
The purpose of this section is to present the determination of stress fields and give full evidence for the important cases when invisibility or stress annihilation occurs, two situations in which the solution does not display a singular behaviour, so that the material is in an optimal situation with respect to failure.

\begin{figure}[!htb]
  \begin{center}
\includegraphics[width=15 cm]{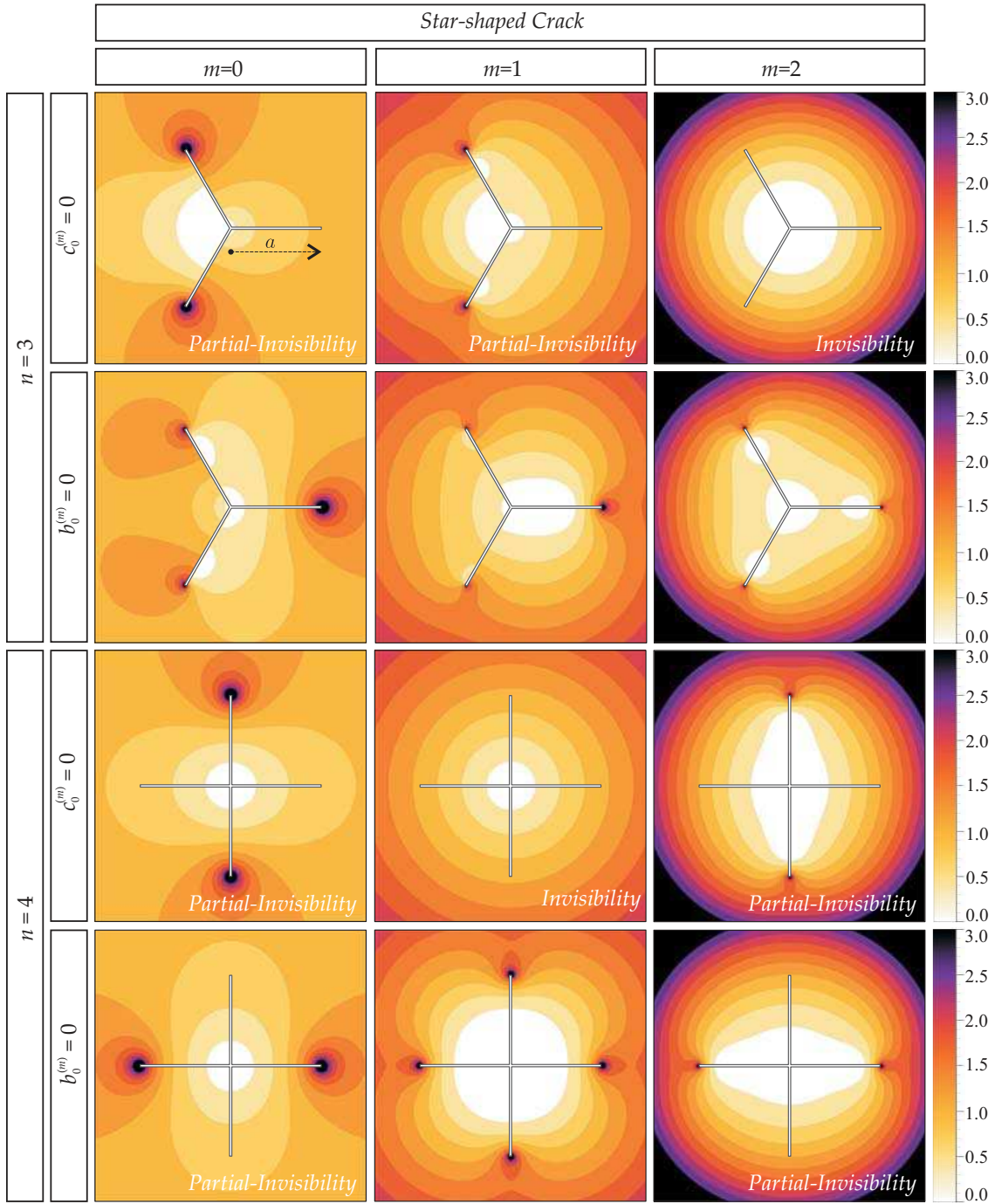}
\caption{
Level sets of shear stress modulus ($\tau^{(m)}(x_1, x_2)/\tau^{\infty(m)}(a,0)$) near $n$-pointed star-shaped cracks ($n=3$ and 4),
for different orders $m$ of the applied remote polynomial antiplane shear loading
($m=0$ uniform, $m=1$ linear, and $m=2$ quadratic shear loading).
Some cases of invisibility and partial-invisibility are shown; note that invisibility  occurs only for certain combinations of $n$ and $m$ (while for the single crack invisibility
is independent of $m$, see Fig. \ref{fig_part_I_crack_fullfield_curve}).}
\label{fig_part_I_34crack_fullfield}
 \end{center}
\end{figure}

\begin{figure}[!htb]
  \begin{center}
\includegraphics[width=15 cm]{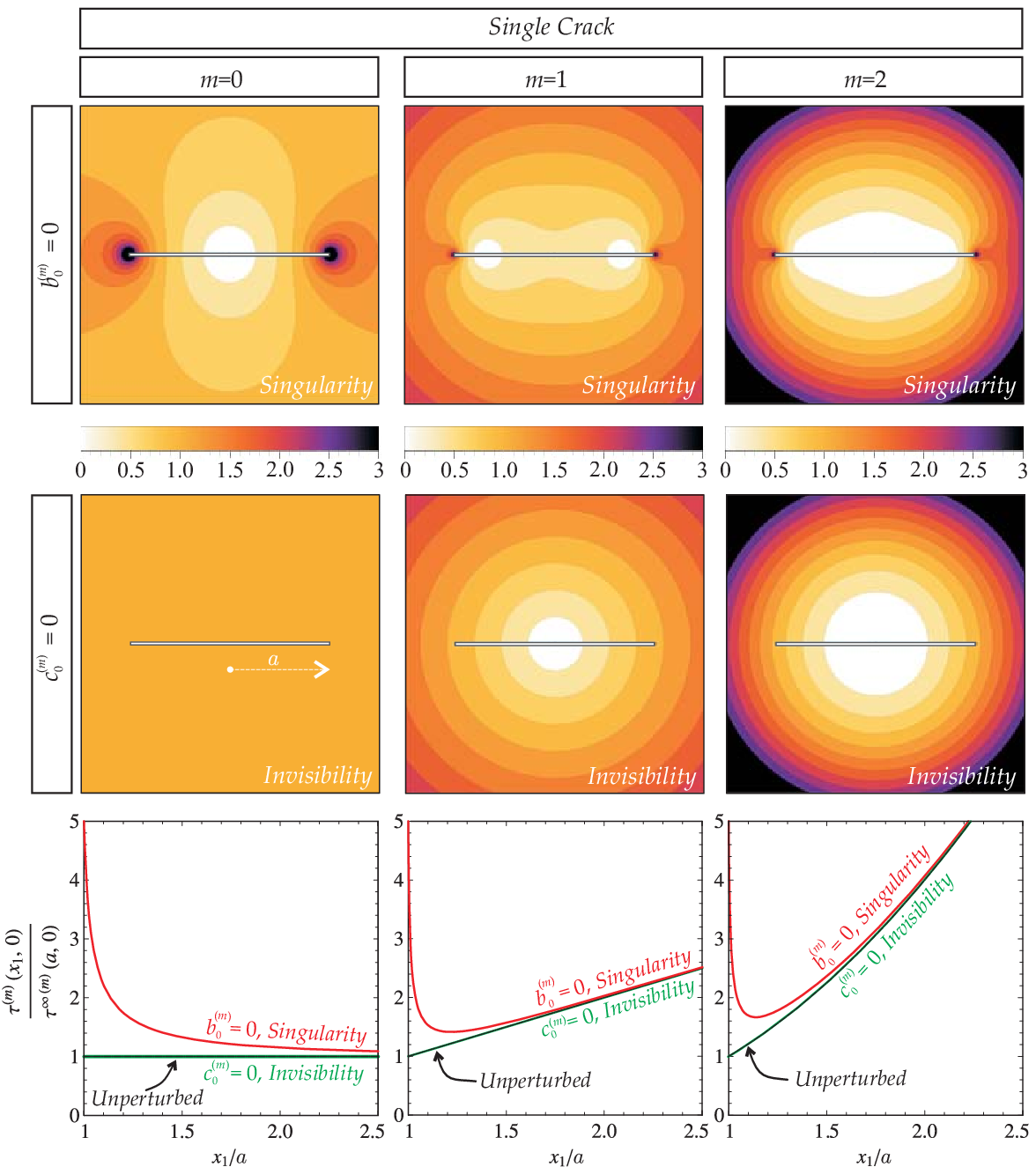}
\caption{Usual cases of stress singularity ($b_{0}^{(m)}=0$, upper part), versus invisibility ($c_{0}^{(m)}=0$, middle part) for a single
crack evidenced by the field of dimensionless modulus of shear stress
($\tau^{(m)}(x_1, x_2)/\tau^{\infty(m)}(a,0)$). In addition, shear stress modulus ($\tau^{(m)}(x_1, 0)/\tau^{\infty(m)}(a, 0)$, lower part)
ahead of the crack tip  is also detailed (lower part).
Different orders $m$ of the applied remote polynomial antiplane shear loading  are considered ($m=0$ uniform, $m=1$ linear, and $m=2$
quadratic shear loading).
The single crack (or the single stiffener), $n=2$, is the only case for which the invisibility
holds at every order of applied remote shear loading, $m=0,1,2,\dots \in \mathbb{N}$.
}
\label{fig_part_I_crack_fullfield_curve}
 \end{center}
\end{figure}

\begin{figure}[!htb]
  \begin{center}
\includegraphics[width=15 cm]{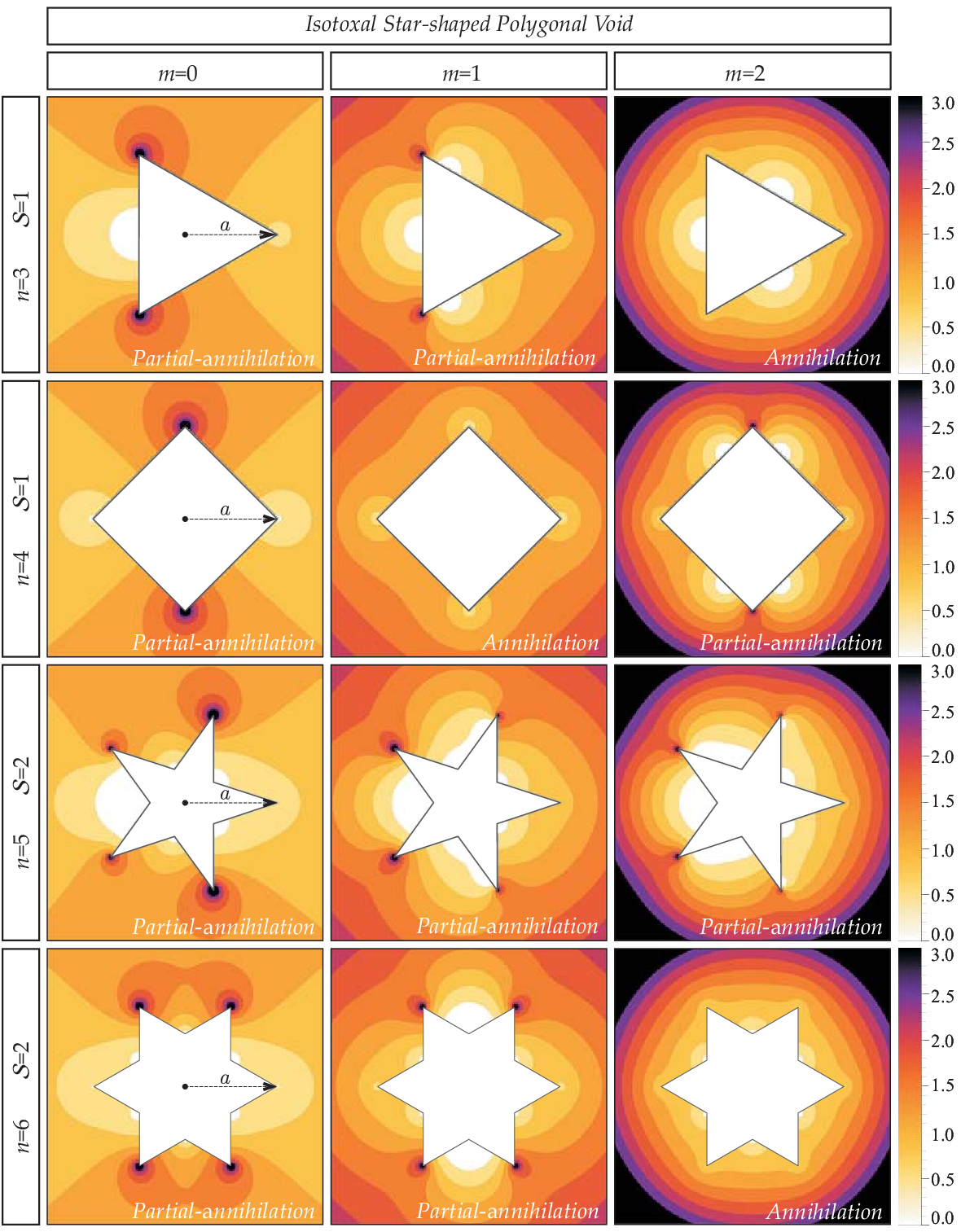}
\caption{
Level sets of shear stress modulus ($\tau^{(m)}(x_1, x_2)/\tau^{\infty(m)}(a,0)$) near $n$-pointed isotoxal star-shaped voids,
for different starriness ${\mathcal S}$ and orders $m$ of the applied remote polynomial antiplane shear loading
($m=0$ uniform, $m=1$ linear, and $m=2$ quadratic shear loading) for $c_0^{(m)}=0$.
 Some cases of stress annihilation and partial-annihilation are shown, which occur only for certain combinations of $n$ and $m$.}
\label{fig_3crack_fullfield}
 \end{center}
\end{figure}

\begin{figure}[!htb]
  \begin{center}
\includegraphics[width=11 cm]{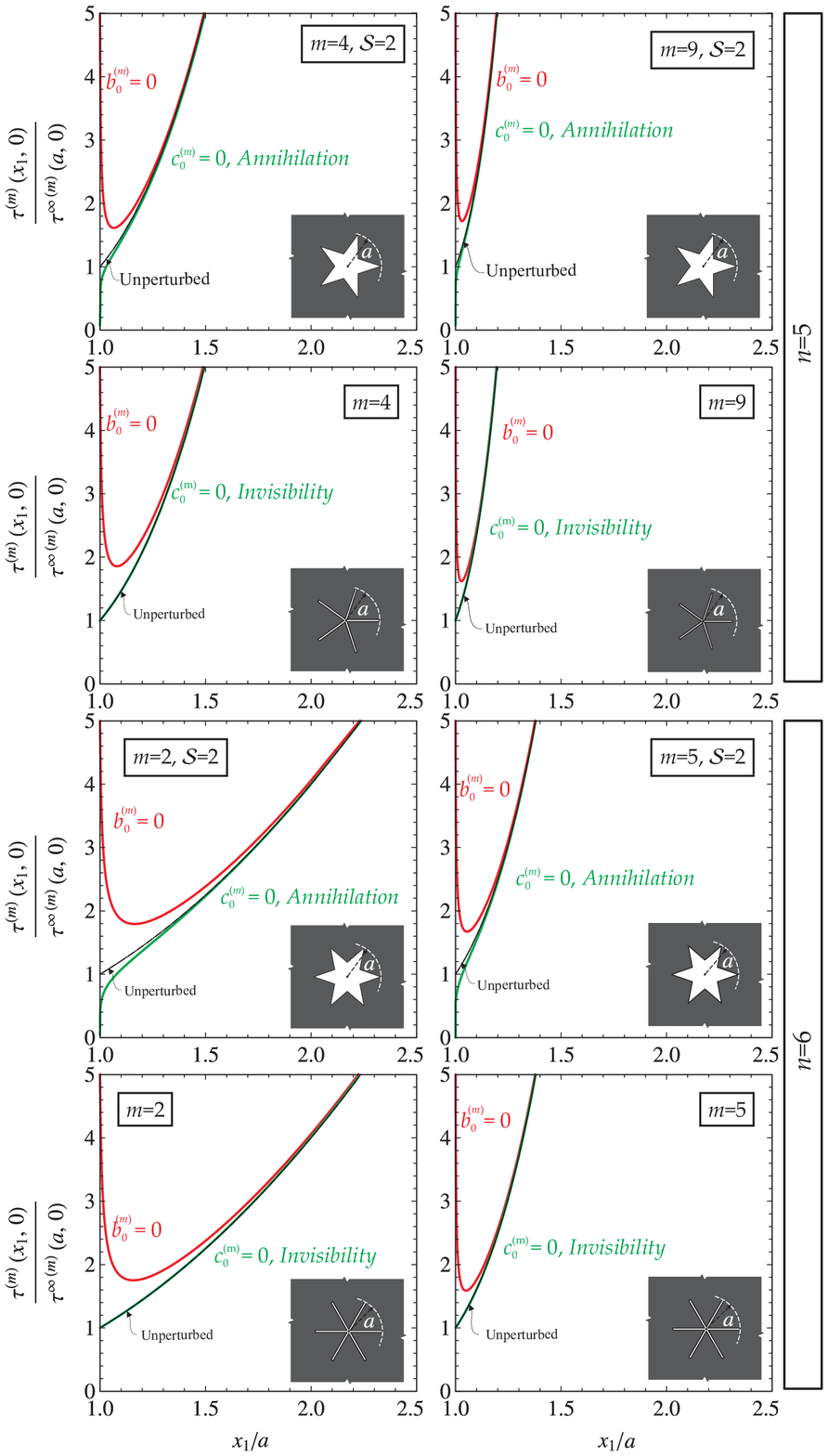}
\caption{Cases of stress annihilation of a star-shaped isotoxal polygonal void and star-shaped crack, for
different values of $m$ and $n$.
The shear stress modulus ahead of one of void vertex ($\tau^{(m)}(x_1, 0)/\tau^{\infty(m)}(a, 0)$) is reported for the cases $b_0^{(m)}=0$ (red), $c_0^{(m)}=0$ (green), and unperturbed (black). Note that the green curve coincides with the black curve when inclusion invisibility occurs.}
\label{fig_part_II_poly_curve}
 \end{center}
\end{figure}

\begin{figure}[!htb]
  \begin{center}
\includegraphics[width=13 cm]{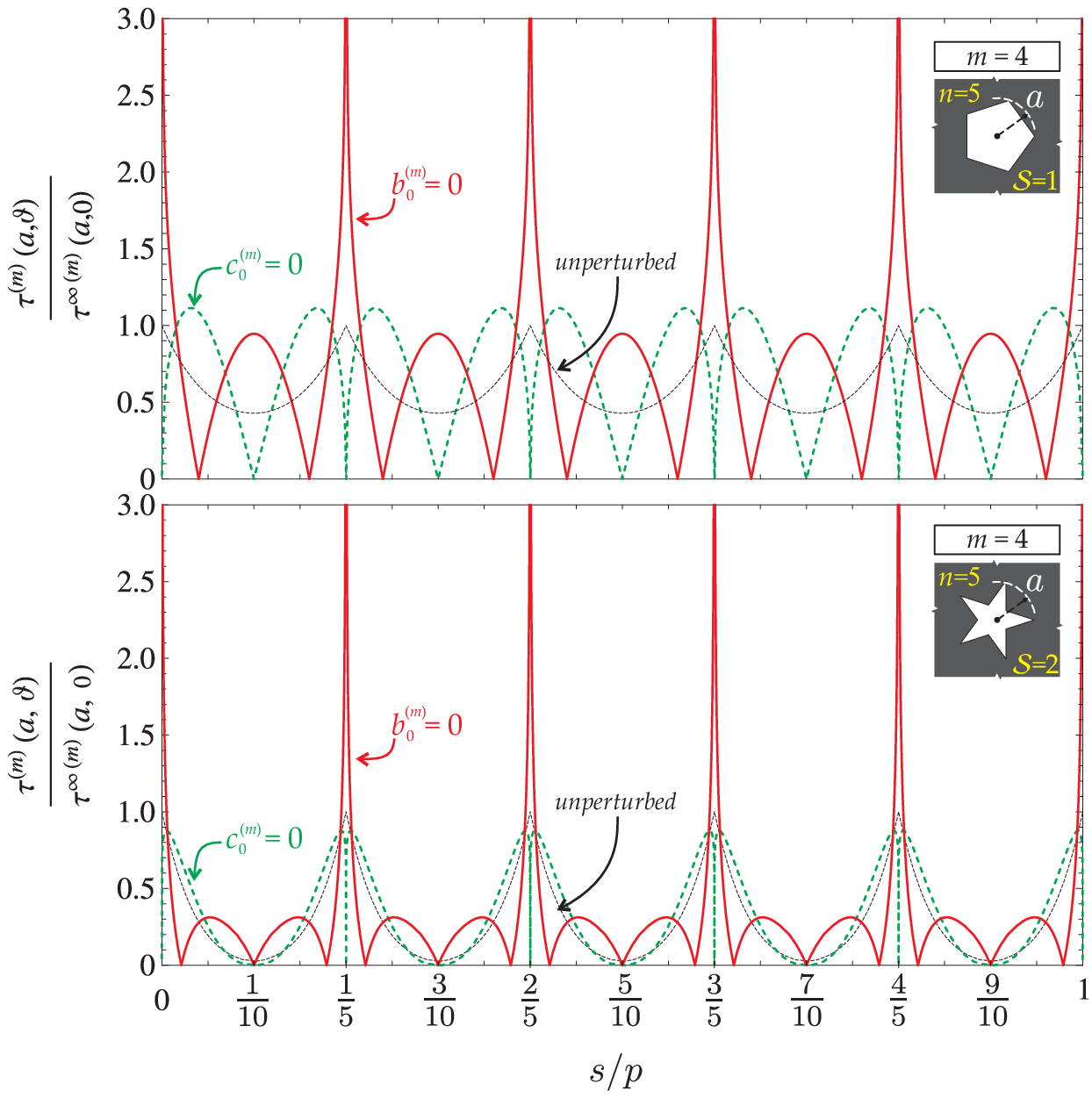}
\caption{ Stress singularity ($b_{0}^{(m)}=0$, reported in red) and stress annihilation ($c_{0}^{(m)}=0$, reported in green) for a regular pentagon-shaped ($n=5$, $\mathcal{S}=1$, upper part) and regular pentagram-shaped
($n=5$, $\mathcal{S}=2$, lower part) void for the case of fourth-order remote shear field $m=4$. The modulus of the shear stress is reported along the perimeter of the figures, measured by the coordinate $s$ (made dimensionless through division by the length of the perimeter $p$), so that
$s/p=j$ corresponds to the j-th vertex.
Note that the singularity produced by the pentagram-shaped void is stronger than that produced by the pentagonal void, because the angles at the vertex of the pentagram are sharper than that of the curvilinear coordinates along the inclusion perimeter the pentagon.}
\label{fig_part_II_curve_lungo_perimetro}
 \end{center}
\end{figure}

\begin{figure}[!htb]
  \begin{center}
\includegraphics[width=12 cm]{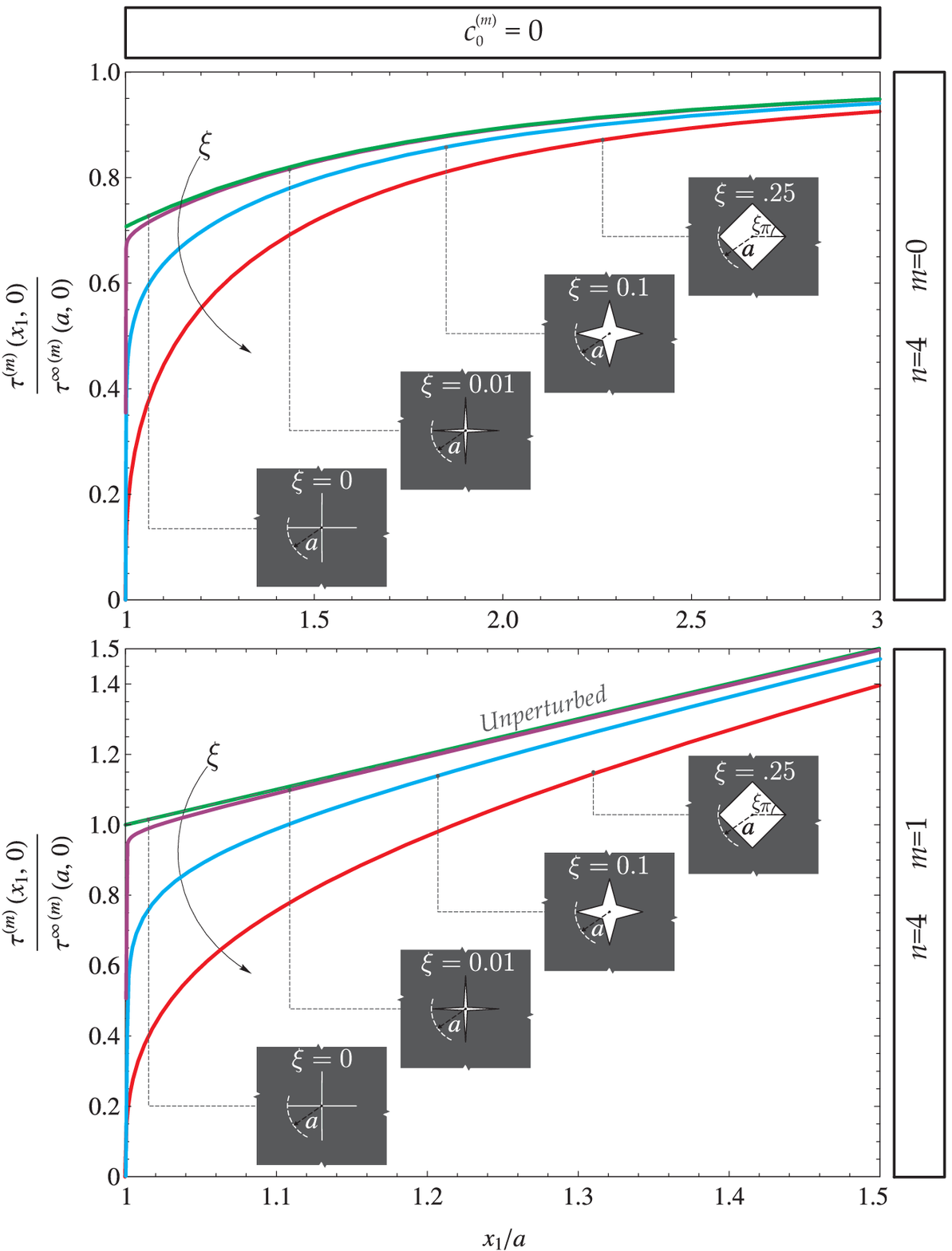}
\caption{Effect of the reduction in the semi-angle of an inclusion vertex ($\xi \pi$) on the modulus of the shear stress ahead of the vertex.
The shear stress distribution tends to that corresponding to a star-shaped crack when
$\xi$ tends to zero (note the boundary layer). A uniform remote stress field ($m=0$, $c_{0}^{(0)}=0$)
is considered on the left for a four-pointed star, while a case corresponding to stress annihilation and invisibility is considered on the right ($m=1$, $c_{0}^{(1)}=0$).
}
\label{fig_part_II.transition}
 \end{center}
\end{figure}

\begin{figure}[!htb]
  \begin{center}
\includegraphics[width=10 cm]{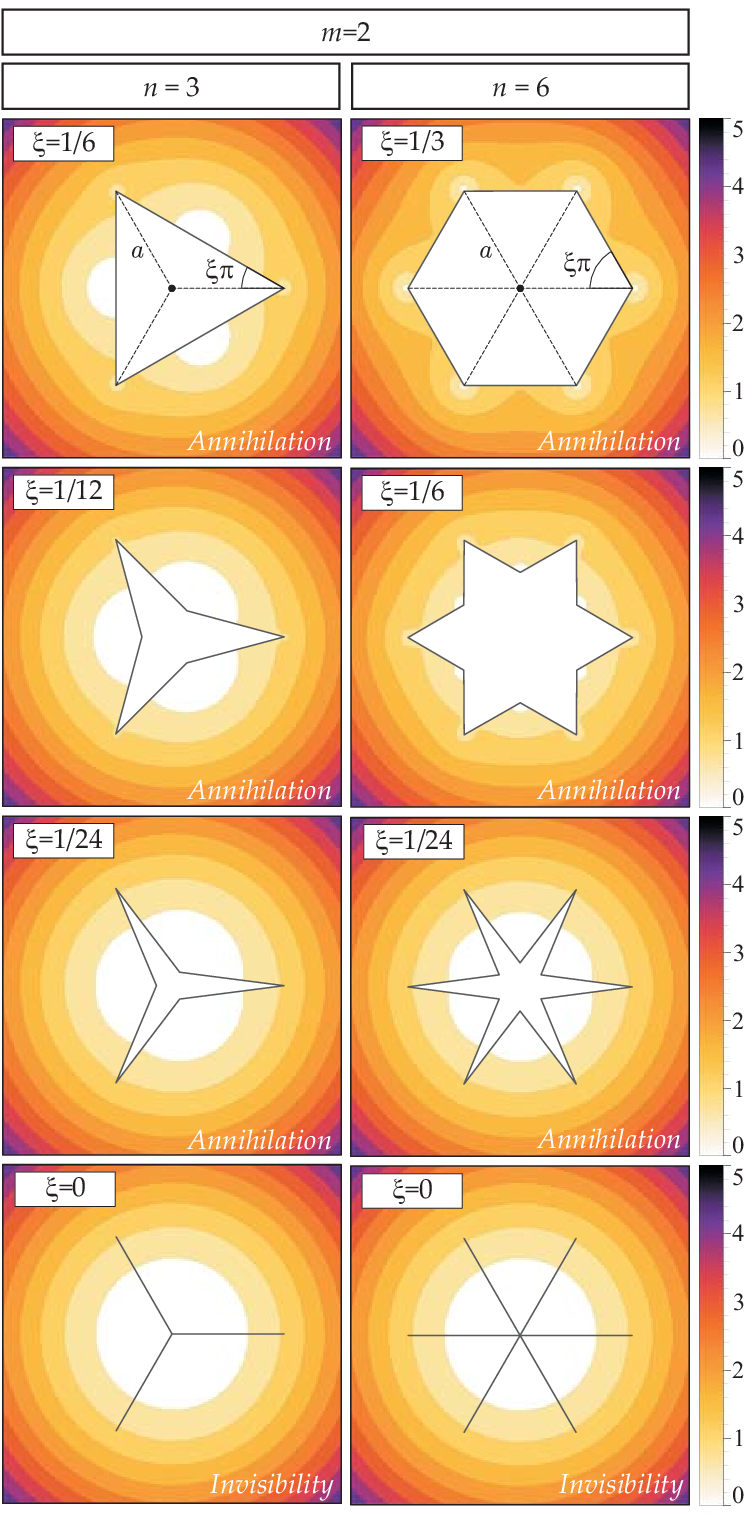}
\caption{Cases of stress annihilation of $n$-pointed isotoxal polygonal voids of star-shaped cracks subject to a quadratic remote antiplane
shear load ($c_{0}^{(m)}=0$) are shown for different values of $\xi$. Level sets of shear stress modulus ($\tau^{(m)}(x_1, x_2)/\tau^{\infty(m)}(a,0)$) are reported, in particular the transition from stress annihilation to the invisibility occurs when $\xi$ tends to zero.}
\label{fig_part_II_transition_fullfield}
 \end{center}
\end{figure}

The stress fields that are presented in Figs. \ref{fig_part_I_34crack_fullfield} -- \ref{fig_part_II_transition_fullfield} refer
for simplicity to cracks and voids, but it is known from Section 3.3 of Part I that a stress analogy holds for a  void and a rigid
inclusion of identical shape, so that the presented results are also valid for rigid inclusions and stiffeners, with the proper interchange
between constants $b_{0}^{(m)}$ and $c_{0}^{(m)}$ as indicated in Section 3.3 of Part I.

Level sets of the modulus of the shear stress are reported in Fig. \ref{fig_part_I_34crack_fullfield} for three and four pointed ($n$=3, 4)
star-shaped cracks with uniform ($m$=0), linear ($m$=1), and quadratic ($m$=2) remote stress field.
Note that \lq normally' the crack strongly perturbs the stress field and induces a square-root singularity at each point of the star.
In special cases ($\{m, n\}=\{2,3\}$ and $\{m, n\}=\{1,4\}$ marked in the figure), however, the solution remains completely
unperturbed (so that the level sets of the modulus of shear stress displays a concentric circular structure). In these cases
the star-shaped crack (or stiffener, with the due changes in the loading coefficients, see Section 3.3 of Part I)
remains \lq completely neutral' or \lq invisible'.

The single crack or stiffener is the only example of inclusion which can remain invisible at every loading order $m$.
This invisibility occurs for $c_0^{(m)}=0$ and is shown (for $m=0,1,2$) in the middle part of Fig. \ref{fig_part_I_crack_fullfield_curve},
while in the upper part the \lq usual' cases of stress singularity are reported, for $b_0^{(m)}=0$.
The situation of a single crack (or stiffener) is also detailed in the lower part of Fig. \ref{fig_part_I_crack_fullfield_curve}, where the modulus of the shear stress is reported ahead of the crack (or stiffener) tip and the
difference between singularity and invisibility can be further appreciated.

Results for isotoxal star-shaped polygonal voids are reported in Fig. \ref{fig_3crack_fullfield} in terms of level sets of the modulus of the shear stress for regular polygonal ($n$=3, 4 and $\mathcal{S}=1$) and regular star-shaped ($n$=5, 6 and $\mathcal{S}=2$) polygonal voids at uniform ($m$=0), linear ($m$=1), and quadratic ($m$=2) remote stress field.

Fig. \ref{fig_part_II_poly_curve} shows  the modulus of the shear stress ahead of the star-shaped polygonal void
($n$=5, 6 and $\mathcal{S}=2$) and star-shaped crack ($n$=5, 6), at different values of $m$ (=2,4,5,9), and the
difference between singularity, annihilation and invisibility can be observed.

Two cases of stress singularity and stress annihilation along the perimeter of a pentagonal void and a pentagram-shaped void are reported in Fig. \ref{fig_part_II_curve_lungo_perimetro}.

An isotoxal star-shaped polygonal void tends to a star-shaped crack when the semi-angle at the inclusion vertex ($ \xi \pi$) decreases and tends to zero. It is therefore expected that the stress field generated near an isotoxal star-shaped void tends to that corresponding to a star-shaped crack.
This tendency is confirmed by the results shown in Fig. \ref{fig_part_II.transition}, where the modulus of the shear stress
ahead of a void vertex for different values of $\xi$ is presented.
Note that in the right part of Fig. \ref{fig_part_II.transition} the conditions have been selected ($n=4$, $m=1$) for which the star-shaped crack becomes invisible and the polygonal void exhibits the stress annihilation.

The same situation of a polygonal and star-shaped void for which the stress annihilation is verified and the parameter $\xi$ is decreased is also reported in Fig. \ref{fig_part_II_transition_fullfield}, where the level sets of the shear stress modulus $\tau^{(m)}(x_1, x_2)/\tau^{\infty(m)}(a,0)$ for $n=3,6$ and $m=2$ are plotted.
It may be concluded from this figure that, for decreasing values of the semi-angle at the inclusion vertex ($\pi \xi$), a transition can be observed from stress annihilation to a sort of \lq quasi-invisibility', obtained in the proximity of the limit of star-shaped crack, where the stress field is only slightly perturbed.

\begin{figure}[!htb]
  \begin{center}
\includegraphics[width=16 cm]{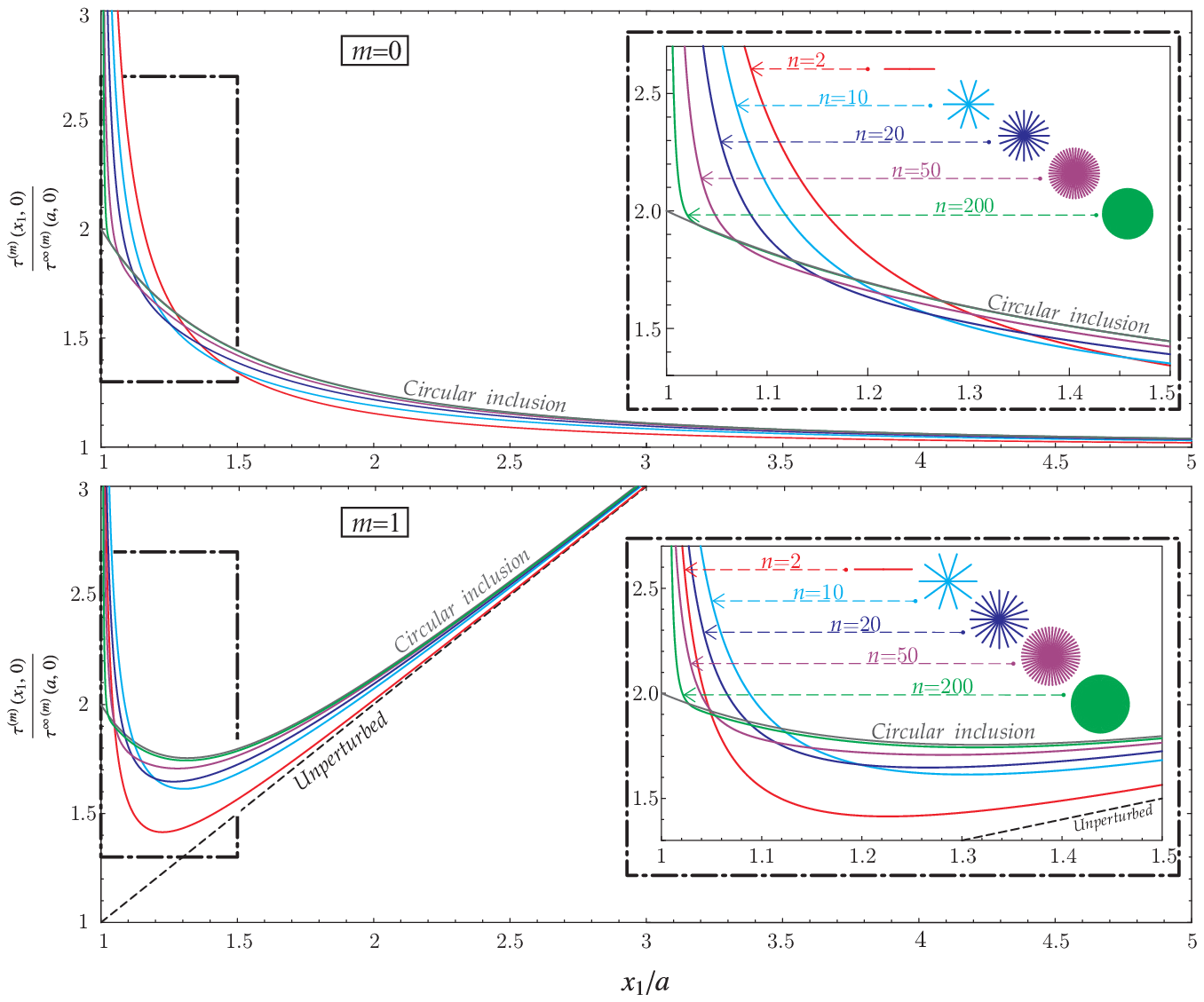}
\caption{Shear stress modulus ($\tau^{(m)}(x_1, 0)/\tau^{\infty(m)}(a, 0)$) ahead of the crack lying on the $x_1$-axis of a star-shaped crack, subject to uniform ($m=0$, upper part) and linear ($m=1$, lower part) antiplane shear. Several values of points $n$ are considered, so that it can be noted that the stress field correctly tends to the solution of a circular void (obtained at $n\rightarrow \infty$), which exhibits the well-known value of stress concentration factor (SCF=2), independent of the loading order $m$.}
\label{fig_part_I_transition}
 \end{center}
\end{figure}

An interesting situation can be envisaged in the limit when the number of points of a star-shaped crack grow and tend to infinity.
This is explored in Fig. \ref{fig_part_I_transition}, where the modulus of shear stress
ahead of the crack tip located on the $x_1$-axis is
reported. The stress field tends with increasing $n$ to the stress distribution corresponding to a circular void with radius equal to the crack length.

\subsection{The rules of invisibility and annihilation}

In the previous figures several cases of star-shaped crack (or stiffener) invisibility and stress annihilation at the points of an isotoxal star-shaped void (or rigid inclusion) have been presented.
It is therefore now convenient to establish the rules governing these two important situations.

The formulae of SIFs and NSIFs (\ref{eq_sif}) and (\ref{eq_nsif}), respectively, highlight that the loading coefficients $b_{0}^{(m)}$ and $c_{0}^{(m)}$ control the stress singularity, which can be deactivated
at the vertex aligned on the $x_1$ axis in the following two cases
\begin{itemize}

\item for a crack or a void $\chi=1$ when $c_{0}^{(m)} = 0$,

\item for a stiffener or a rigid inclusion $\chi=-1$ when $b_{0}^{(m)} = 0$.

\end{itemize}

The stress singularity can simultaneously be deactivated at all vertices of the void or rigid inclusion, when the above conditions are verified together with the following constraints
between the loading order $m$ and number of the star points $n$
         \beq
                \label{eq_condizioni}
       m=\begin{cases}
        \barr{cll}
             n\, j -1 \qquad \text{if $n$ is odd},\\[6mm]
             \dfrac{n}{2}j-1 \qquad \text{if $n$ is even} ,
                \earr
        \end{cases}
    \eeq
where $j \in \mathbb{N}_1$. Note that eqn (\ref{eq_condizioni}) is equivalent to $2(m+1)/n \in {\mathbb N}_1$.

In the case of star-shaped cracks or stiffeners, conditions (\ref{eq_condizioni}) imply that the perturbed potential, eqn (52)$_2$ of Part I, is pointwise null $g^p=0$, so that the deactivation of the
singularity corresponds to \lq invisibility' or \lq full neutrality'.
In the case of isotoxal star-shaped voids or rigid inclusions, setting $\gamma=2\pi/n$ in eqns (10) of Part I,
conditions (\ref{eq_condizioni}) imply that the loading coefficients satisfy the conditions
$\hat{b}_0^{(m)}=b_0^{(m)}$ and $\hat{c}_0^{(m)}=c_0^{(m)}$, so that the deactivation of the
singularity corresponds to \lq stress annihilation'.
Note that the stress annihilation follows not only from the lack of stress singularity, but also from the fact that the leading-order term in the stress asymptotic expansion is a positive power of the radial distance from the
vertex (Section 2.2 of Part I).

\subsection{Partial-invisibility and Partial-annihilation}

Partial invisibility or partial stress annihilation occurs respectively when some of the crack/stiffener tips are invisible or when some of the points of the star-shaped void/inclusion are stress-free.
In these situations, which are more frequent than the cases of full annihilation and invisibility, the material fails at one of the points where the stress remains singular.

Cases of partial invisibility are reported in Fig. \ref{fig_part_I_34crack_fullfield}. In particular, the horizontal crack is invisible  for $n=3$, $m=1$, and $c_0^{(m)}=0$ and for $n=4$, $m=0, 2$, and $c_0^{(m)}=0$;
the vertical crack is invisible for $n=4$, $m=0, 2$, and $b_0^{(m)}=0$.
All examples reported in Fig. \ref{fig_3crack_fullfield} present at least partial stress annihilation, which occurs at all vertices located along the horizontal $x_1$-axis.

\subsection{Beyond the hypotheses of infinite matrix and regular shape inclusion} \lb{sonostanco}

Inclusion invisibility and stress annihilation have been demonstrated under the assumptions that the inclusions have a {\it regular} shape and that the matrix is {\it infinite}.
It is suggested in this Section that these two hypotheses can be relaxed, so that the two concepts of invisibility and stress annihilation are more general than it has been demonstrated analytically.

\begin{figure}[!htb]
  \begin{center}
\includegraphics[width=12 cm]{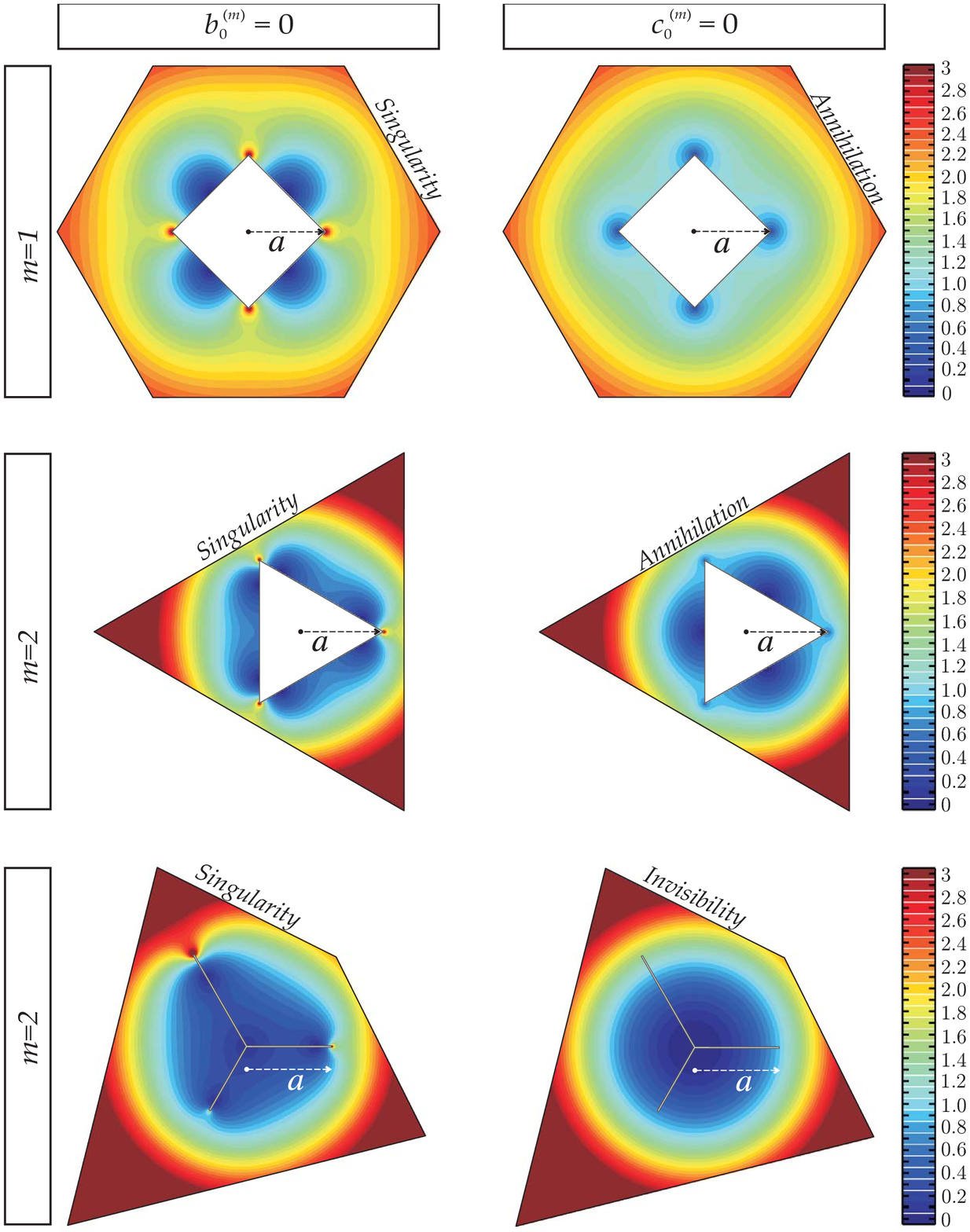}
\caption{Effect of finiteness of the domain in which a polygonal void is embedded, evidenced through the level sets of shear stress
modulus $\tau^{(m)}(x_1, x_2)$ (made dimensionless through division by $\tau^{\infty(m)}(a, 0)$)
 obtained numerically for three different geometries:
(upper part) a regular hexagonal matrix containing a concentric square void, (central part) a regular triangular matrix containing a concentric
triangular void, and (lower part) a quadrilateral containing a three-pointed irregular star-shaped crack.
Stress singularity is obtained for $b_0^{(m)}=0$ (left),
while stress annihilation (for star-shaped polygonal void) and invisibility (for star-shaped crack) occurs for $c_0^{(m)}=0$ (right).}
\label{ultimafig}
 \end{center}
\end{figure}

A finite portion of an elastic body is considered ideally \lq cut' from the infinite elastic space subjected to the $m$-th order polynomial field considered in Part I of this study [equations (11) and (12)]. Therefore, the boundary of the cut body is assumed subject to the traction conditions transmitted by the rest of the elastic space.
In this way, polynomial displacement and stress fields are realized within the body in the absence of any inclusion. When invisibility is achieved for the infinite body (as related to the presence of zero-traction or zero-displacements lines,  inclined at $\widehat{\theta}_j=j \pi/(m+1)$, $j\in [0,...,2m+1]$), it will also hold for the body \lq cut' from the infinite body. This is shown in Fig. \ref{ultimafig} (lower part), where invisibility is achieved for a
finite elastic body and irregular star-shaped crack.

If now the cut body has an external boundary with the shape of a regular polygonal shape, symmetric with respect to all the inclinations
$\widehat{\theta}_j$, the introduction of a $n$-sided regular polygonal void (or a rigid inclusion) satisfying eqn (\ref{eq_condizioni}),
realizes a stress field which annihilates at the inclusion vertices, as Fig. \ref{ultimafig} (upper and central parts) shows for polygonal voids.

The results presented in Fig. \ref{ultimafig}, in terms of modulus of shear stress level sets near polygonal voids,
have been obtained numerically (using the finite element software Comsol Multiphysics$^\copyright$ version 4.2a,
by exploiting the analogy between the antiplane problem and the 2-dimensional heat transfer equation under stationary conditions), by imposing
traction boundary conditions at the external contour of the body.
To avoid the presence of a rigid-body motion, the open boundary condition available in the software has been selected.

In the simulations, the domains have been discretized at two levels
by the free triangular user-controlled custom mesh. At the
first level, the entire domains are meshed with maximum and
minimum element sizes equal to $10^{-2}R$ and $10^{-5}R$, respectively,
where $R$ denotes the unit radius of the circle inscribing the
domain. At the second level, the inclusion boundaries have been meshed with maximum and minimum element
sizes equal to $10^{-3}R$ and $10^{-5}R$, respectively.

\section{Conclusions}

The closed-form expression for the SIFs and the NSIFs have been obtained
for isotoxal star-shaped polygonal voids or rigid inclusions (including the limits of star-shaped cracks or stiffeners)
loaded in an infinite elastic plane through a nonuniform antiplane shear loading.
Two special cases have been identified in which (i.)
the stress is annihilated at all of the vertices of a star-shaped void or rigid inclusion,
and (ii.) a star-shaped crack or stiffener becomes quasi-statically invisible, leaving the ambient field
completely unperturbed.
These situations are of particular interest because in these cases the failure of the material does not occur at the vertices of the inclusions.
Therefore, the results presented in this paper describe optimal loading conditions that can be used in the design of novel composite materials.

\section*{Acknowledgments}
The authors gratefully acknowledge financial support from the ERC Advanced Grant \lq Instabilities and nonlocal multiscale modelling of materials'
ERC-2013-ADG-340561-INSTABILITIES  (2014-2019).

\end{document}